# Exploring the context of visual information seeking

**Shahram Sedghi**

Department of Library and Medical Information Sciences, Iran University of Medical Sciences, Tehran, Iran

**Zeinab Shormeij**

Department of Library and Medical Information Sciences, Iran University of Medical Sciences, Tehran, Iran,

**Iman Tahamtan**

School of Information Sciences, College of Communication and Information, University of Tennessee, Knoxville, Tennessee, USA

## Abstract

**Purpose** – Information seeking is an interactive behaviour of the end users with information systems, which occurs in a real environment known as context. Context affects information-seeking behaviour in many different ways. The purpose of this paper is to investigate the factors that potentially constitute the context of visual information seeking.

**Design/methodology/approach**: We used a Straussian version of grounded theory, a qualitative approach, to conduct the study. Using a purposive sampling method, 28 subjects participated in the study. The data were analysed using open, axial and selective coding in MAXQDA software.

**Findings**: The contextual factors influencing visual information seeking were classified into seven categories, including: "user characteristics", "general search features", "visual search features", "display of results", "accessibility of results", "task type" and "environmental factors".

**Practical/implications**: This study contributes to a better understanding of how people conduct searches in and interact with visual search interfaces. Results have important implications for the designers of information retrieval systems.

**Originality/value**: This paper is among the pioneer studies investigating contextual factors influencing information seeking in visual information retrieval systems.

**Introduction**

Understanding context is important for obtaining better knowledge of individual information activities (Starasts, 2015). Thus, it is important to know what elements create the context, how context is understood (Courtright, 2007) and how the information process relates to its context (Kari and Savolainen, 2007). Nevertheless, there is not a strong consensus on what constitutes a context, and what relationships exist between individuals' information seeking and its context (Chang and Lee, 2001). The importance of context in information-seeking behaviour and the information retrieval (IR) (Kelly, 2006) as well as the importance of understanding information seeking and retrieval within different contexts has been emphasised in a large number of previous studies as Albertson (2015) reported. However, to the best of our knowledge, there is a lack of knowledge regarding how users interact with visual search interfaces. Visual search interfaces are emerging as a new way to help users respond to their information needs effectively. For this reason several library vendors such as EBSCOhost have added visual search interfaces or functions to their products (Fagan, 2006). Current study aims to find the factors that potentially constitute the context of visual information seeking.

**Literature review**

*Information seeking and context*

Information seeking is a dynamic and interactional process, which occurs in a real environment referred to as the context (Mai et al., 2016). The context involves the information seeking of individuals formed in interaction with other people, social networks and various situations, tools and so on (Wilson, 1997). Dervin (2003, p. 112) noted that "there is no term that is more often used, less often defined, and when defined so variously as context". In information-seeking research, "context" describes the situation surrounding a person's information-seeking behaviour (Dervin, 2003), and a setting where information activities take place. In addition, context is defined as, "any factors or variables that are seen to affect individuals' information-seeking behaviour: socio-economic conditions, work roles, tasks, problem situations, communities and organisations with their structures and cultures, etc" (Talja et al., 1999, p. 752).

Context consists of several elements, each of which is defined by several contextual factors. In this regard, Kari and Savolainen (2007) listed three relationships between context and information behaviour: association relationship, interaction relationship and one-directional relationship.

Association relationship is defined as the concurrence of certain information behaviour(s) in a situation. Interaction implies that the context first influences the individual's information behaviour, and then this behaviour influences contextual factors. The one-directional relationship refers to the contextual factors that encourage, affect, determine or prevent some information behaviours of the individual.

Several previous studies and theoretical models have examined the relationships among contextual factors and information activities (Courtright, 2007). Kelly (2006) reviewed some of the factors that potentially constitute information-seeking context. For instance, organizational culture, task (Courtright, 2007), time (Savolainen, 2006), domain knowledge (Wildemuth, 2004), attitude, cognitive factors, computer anxiety, discipline, expertise, gender, meta-cognition and social factors (Kari and Savolainen, 2007) are mentioned as contextual shaping factors in information-seeking behaviour.

*Visual information retrieval systems*

In a number of studies, researchers have attempted to create more effective tools and methods, such as visual IR systems (Heilig et al., 2008), to facilitate user information seeking by providing them with interfaces that are user friendly and have a satisfying usability (Gerken et al., 2009).

However, as information is rapidly growing in quantity, heterogeneity and dimensionality, designing such systems has become an increasingly difficult task (Gerken et al., 2009). As the interaction between users and visual representations takes place during information seeking, a poor representation could not only burden users more but also disturb their information-seeking process (Song, 2000).

The literature suggests that integrating information visualisation (IV) and IR yields several advantages for information seekers. For instance, "visualization-based search engines improve the efficacy and accuracy of IR, particularly when common, non-specific queries are used" (Pajić, 2014, P. 146). Individuals who are less familiar with specialized terminologies benefit the most from visual IR systems (Wu et al., 2008). Visual representations of retrieved documents (e.g. bar charts for numerical values) and/or keywords in the form of (concept) maps facilitate better cognitive fit for human information processing and enhance the users' search efficiency (Huang et al., 2006; Shen et al., 2006)

EBSCOhost is now one of the largest scientific information providers of full text and bibliographic databases designed for research and provides users with visual search options. In the visual representation of EBSCOhost, the main categories (root) were linked to a collection of more specific items. In other words, each item used to have a link to one parent item. According to Paji_c (2014, p. 146):

*EBSCO Visual Search used to create and visualize clusters of documents. Single documents were displayed as rectangles, while clusters of related documents were represented as circles. The color of the object indicated document's age (blue-old, red-new).*

In the later version, EBSCO simplified its visualisation by using a hierarchical order of documents, which were presented as a tree of subheadings related to the terms in the initial query (Pajic, 2014).

Many studies have considered the need for comprehending the reality of contextual factors for improving user interaction with search. However, the contextual factors influencing information seeking in visual IR systems have not yet been well studied in the literature.

**Methodology**

We used a qualitative research methodology to investigate the contextual factors affecting users' interaction with the visual search of EBSCO. Qualitative studies provide a broad picture of information seeking and enable researchers to understand major contextual factors in information seeking (Starasts, 2015). Most studies on the topic of context are based on qualitative research and have their data gathered through interviews, because exploratory studies provide a deeper insight into information-seeking behaviour and enable researchers to obtain a more in-depth understanding of the reality of the context.

Qualitative research is appropriate for digging deep into the contextual factors influencing information seeking behavior (Kari and Savolainen, 2007). In qualitative studies, recruiting samples carries on until data saturation is achieved. That is to say, data collection is continued, until the data collected from participants becomes repetitive and no longer adds to the collected data and data saturation occurs (Mason, 2010). In general, in qualitative studies, sample size is a relative matter that cannot be too small or too large. Given these points, 28 interviews were conducted in the present study.

Using a purposive sampling method, the respondents were selected from two groups at a medical university: Group A included 20 graduate students (PhD and master's levels) of various disciplines, and Group B included eight medical librarians. The reason for choosing two different groups was to obtain distinct perspectives on the issue under study and to ensure the interviewees reflect these differences (Rubin and Rubin, 2011). Graduate students had a proper knowledge of information seeking in scientific databases. All librarians who participated in the study were experienced in searching in search engines and scientific databases such as EBSCO.

Participants were presented with a general description of EBSCOhost and its features. No time limit was imposed and subjects were asked to search a topic of their own interest in the database until they achieved their required results. Semi-structured interviews were used with open-ended questions. In the semi-structured interviews, "the interviewer starts with pre-planned questions and then probes the interviewee to say more until no new relevant information is forthcoming" (Rogers et al., 2011, p. 230). When the interviewer asks an open question, there would be no expectation in terms of the format, content or length of the answers (Rogers et al., 2011). Thus, in this study, the interviewees were free to answer – or not – the questions, describe, express and clarify their responses.

As the best time for an interview varies across individuals (Rubin and Rubin, 2011), participants were interviewed when their time was not occupied. Decisions about where to conduct interviews were made by participants to meet the needs of comfort and confidentiality (Rubin and Rubin, 2011). Thus, most interviews took place at the user's office and at their preferred day and time. During the interviews, the user's information-seeking behaviours were recorded using Camtasia software to enable researchers to analyse the collected data more accurately, and, if necessary, to match the users' statements with their recorded videos. The interviews lasted from 20 to 60 min based on the respondent's degree of willingness.

Informed consent was taken from participants verbally. Participants were assured that the data were to remain confidential, which insulates their opinions from being seen by others. This insulation helps to discover people's private thoughts, opinions and the things they prefer to keep from other people, which probably are what researchers want to know (Jessor, 1996).

Grounded theory (GT) was adopted to analyze the data. More specifically, the Straussian version of GT was used, as it is very structured. The analytical procedures from Straussian GT include three stages: open, axial and selective coding. MAXQDA, a qualitative data analysis software, was used to speed up the coding process. In open coding, the researchers read the transcripts, extracted and coded concepts and sorted the list of codes based on their semantic similarity. In axial coding, the relationships between similar concepts are examined, and the codes and analytical notes with similar meanings were categorised. In selective coding, further coding of newly emerged concepts was done to create the final themes. Complementary opinions of two experts in the field of information sciences were also obtained about the themes to modify them if necessary. In addition, themes and descriptions were checked with participants to determine whether participants felt they were accurate. Checking results with experts and participants helped to validate the research findings (Lindlof and Taylor, 2011).

**Findings**

Factors affecting interaction with visual search were grouped into several general categories. These factors were, in fact, the main contextual concepts by which the users were influenced in different situations. Some themes were complex and had intricate relationships with one another, which could not be classified in a category with distinctive qualities.

The general categories and core themes developed in the present research included "user characteristics", "general search features", "visual search features", "display of results", "accessibility of results", "task type" and "environmental factors".

*Category 1: Users' characteristics*

Certain characteristics, attributes, and capabilities of users affected how they interacted with the visual search interface. The general characteristics of users affecting the search process were classified into six sub-categories:

*(1) Degree of information need*: Participants had different degrees of information. Interviews revealed that individuals' inclination toward searching was based on the degree of their information need in line with the intended goal. Users with a high degree of information need usually spent more time on completing the task, viewed more documents and used more search tools in EBSCO, such as results refinement. For instance, a first-year master's student initially appeared to have a low degree of information need in comparison to a PhD candidate who was in the middle of writing his dissertation.

As a result, the PhD student dedicated more time and effort into searching in EBSCO than the master's student.

*(2) Domain knowledge*: Domain knowledge is a searcher's familiarity with a subject area, as well as the experts, relevant journals, conferences and organisations in that area. We found that the users' level of knowledge about the subject predicted their success in retrieving the required documents. Users with greater knowledge (mainly PhD students) could easily frequently identified and chose the main concepts pertinent to their topic of interest.

*(3) Search habits (preferences)*: Interviewees asserted that they could have established a better interaction with the search interfaces that they were already familiar with, such as PubMed or ScienceDirect. Many participants preferred other databases regardless of the strengths of the visual search. This raised the insight that users' previous search habits may influence their current or future search behaviours. Thus, being accustomed to a search interface, affects the simplicity or difficulty of a user's interaction with the interface. For example one of the participants said: "As I am used to basic search interfaces, no matter how much easier using this visual interface is [. . .] I prefer basic search over the visual interface. It would be very difficult".

*(4) Patience and perseverance*: Personal and mental attributes, such as a user's patience and perseverance in the search process were among the criteria that affected their information seeking. We found that using the visual information system for finding a document was a time-consuming process, which required users to be patient. Patient users with higher patience and perseverance level chose more terms and sub-terms, opened more documents, and spent more time to achieve their goals.

*(5) Visual search experience*: A participant's familiarity with visual search features could predict the optimal usage of the various facilities of the database and their success in the search process. Participants (mainly librarians) with a higher level of familiarity with visual search could meet their information needs easier and faster.

*(6) Requisite skills*: This refers to skills, such as knowing a particular language (e.g. English) or tools that were required to conduct a successful search. Users with higher-level search who mentioned they were already familiar with the search tools in EBSCO performed more efficient searches, were more successful in fulfilling their information needs, and mentioned to have experienced more positive interactions with the visual search interface.

*Category 2: General search features*

*(1) Search tools*: Users used certain functions of EBSCO, such as refinements, "AND" or "OR" operators, as well as broadening or narrowing search terms to expand or limit the results. Users often began their search with general keywords and, if necessary, used quotation marks and Boolean operators or specific keywords to narrow down their search. In complex searches, users used more diverse terms, more operators and more facilities of the database, to find their required information.

*(2) Other tools and resources*: These tools consisted of "using training manuals" and "starting out the search in Google". User knowledge of tools provided by EBSCOhost could potentially speed up the process of obtaining the desired information and relevant results. Similar to other search engines and databases, EBSCO allocates a section to their training manual, through which users could learn its facilities, specifically the features of the visual search. In addition, some users started their search in Google/Google Scholar, before using EBSCO to obtain information about the topic. These tools and resources helped users to do more successful searches in the database. One of the librarians said: "I mostly use Google and Google Scholar on a daily basis, before searching in a database, searching in databases requires mastery on the search features of the database".

*Category 3: Visual search features*

*(1) Glossary (benefits)*: The glossary of EBSCO helped participants choose proper queries for their searches. The benefits of this glossary included: direct access to the search phrase in the glossary, a hierarchical (tree-shaped) view of the glossary, automatic conversion of the search term to the glossary terms, narrowing down the subject based on the glossary and proceeding step-by-step. A participant said :"It guides users according to what is available, so that if they wanna narrow down their search, they know which branch to follow in order to access the desired results. EBSCO has not gone overboard with anything, and this helps".

*(2) Glossary (deficiencies)*: EBSCO's glossary did not allow outside-the-box thinking and prevented users from creativity and flexibility in their search. Another limitation was the overlap between glossary items so that some topic appeared under several terms. In cases when users were looking for a specific term, they were likely to get confused and not achieve their desired results. In addition, EBSCO had a limited number of unconventional glossary items, which appeared to cause the users to stray from the search path. The following sentence illustrates this limitation: "In health management, I expected to see more options. I went to the health management topic. When [I] narrowed them down, a bunch of terms came up, but I didn't get access to all topics that I had in mind".

*(3) Customizing features*: Like many other scientific databases, EBSCO allowed users to customize their search results. In the visual search, users could refined their search results to the most current documents within a specific time period. Users could also create a folder and save their search results in the folder for future uses. "I chose the subject of medical errors in the visual search platform. The subject was separately classified, and results were narrowed down, so that I could finally get access to the article I wanted, which I could store in a folder for future use", said one of the participants.

*Category 4: Display of search results*

Most databases provide the possibility of displaying retrieved results in a brief format, and EBSCO is no exception. However, its display of search results was different from other databases. Search results were displayed in light and dark fonts, in green and blue colours and in the form of a flowchart in one page. Users could assess the relationship between the retrieved materials and choose relevant records. It also provided the users with the option of displaying the search history, the search strategy and the results, which were accessible in a line view. The following codes were extracted from the interviews:

Displaying retrieved data in light and dark fonts and in green and blue colours on a page: "Yes, I do like the interface of EBSCO, specifically the two colours, and the flowchart format. However, I expected more features from its visual search, and I thought its visual could be nicer".

_ Presenting the display features, such as graphs, diagrams, flowcharts and arrows.

_ Changing the default display of records (column and line view).

_ Automatic display of the search strategy.

_ An entertaining record display environment: "Its visual format is perhaps a kind of game and the best thing about that is its uniformity, through which all the information you need is offered to you in one place".

*Category 5: Accessibility of results*

This theme presents the pros and cons of interaction with the visual search, which is made up of the following subcategories:

Uncertainty about the number of retrieved relevant documents: Given that EBSCOhost visual search only yielded a maximum of 250 results, it left the users unsatisfied with the amount of information retrieved and in general, the search process. In other words, by showing a maximum of 250 results, a proper identification of the number of available material was not possible for the users. "Well, for the limited number of results, we wouldn't know the real number of results, as there's a 250 maximum limit for them, we don't know how many are narrowed down. A search performed in other databases or information systems sometimes gives approximately 10,000 results, and you define your search with the number of results you require. But here, you don't know how many results exist".

*Sorting results according to relevance*: A feature provided by the visual search was the square-shaped icons for presenting the most relevant articles. Some participants mentioned they liked this feature which improved their interaction with the visual interface.

*Limited access to the full-text of articles*: Participants considered access to the full texts of the articles an ultimate success for their searches. In cases, they found articles which were not available via the institution's subscription. "The biggest obstacle is spending a lot of time on a search and finding a paper which is accessible only by purchasing", said a graduate student.

*System responsiveness*: Low speed of accessing results was a major problem for some users. The visual effects of the visual search interface resulted in lowering the speed of accessing the result page for some participants. Contrary to the basic search, retrieving a document in a visual search usually was time-consuming, as it required more than one-time clicking on terms (roots) and sub-terms until the user observed and accessed an article.

Category 6: Task type

We classified each task to "duty-related task" and "curiosity-related task". Duty task is a task for accomplishing specific requirements, according to predefined expectations, which should be done by a specific deadline. Conducting a search in response to others' requests or a course requirement are examples of duty tasks. However, users sometimes merely conduct a search to satisfy their curiosity or to obtain more information about something (curiosity task). In a curiosity task, the user usually has a high degree of information need, which should be done before a specified deadline. The following sentence from one of the interviewees illustrates "the sense of curiosity" for performing a search:

Sometimes neither [do] I have an information need nor a deadline to do a search for a course assignment; rather, I use EBSCO just for my curiosity, probably to find an article which might be or not be in line with my research interests.

Category 7 Environmental factors

From the perspective of some users, environmental factors, such as a calm and proper search situation influenced their interactions with the visual search interface. A quite environment is the minimum requirement for beginning a proper and successful search. Findings indicated that being influenced by environmental factors was a user-dependent issue, while for some it was not too important.

**Discussion**

The value of this study lies in its promotion of understanding information seeking in a visual search interface more contextually, which is significant in improving the optimal understanding of the search task. User characteristics, general search features, visual search features, display of results, accessibility of results, task type and environmental factors were among the main contextual factors affecting users' information seeking in the visual IR system.

Task type is a widely defined contextual factor with significant effects on user interactions with search systems and has been studied extensively (Liu et al., 2010). In the current study, task types were defined as "duty-related task" and "curiosity-related task". A user's success or failure in finding answers to their information may depend on many factors among which is dependent on the task type. Users with a duty task, which is a more complex task, and users with a high degree of information needs dedicate more time and effort to their information seeking, view more documents and show more patience and perseverance in completing the task. Kellar et al. (2007) indicated that for doing information-gathering tasks, which are more complex than many other task types, more pages should be viewed and more time should be dedicated.

In some studies, task complexity is categorized into objective or subjective and its values in both can be low, moderate or high (Liu et al., 2010). Objective task complexity is defined by the number of activities (Ingwersen and Järvelin, 2006) or information sources (Liu et al., 2010) involved in the task. Subjective task complexity is assessed by the task doer in information seeking (Liu et al., 2010), that is, the searcher. According to the present study, if a user has a low degree of information need, he/she might assess the search task to be more complex than a user who has a high degree of information need. The latter is more likely to be more patient during the search task, because what matters for her/him is meeting her/his information need. In addition, results showed that participants used more or less diverse terms, operators and other search facilities of EBSCO to complete the task based on the search difficulty. Aula et al. (2010) found that users formulated more diverse queries, used more advanced operators and spent longer time in difficult tasks. Some studies have also demonstrated that task complexity may affect the type of information needed and the actions taken by the user (Kelly, 2006).

Domain knowledge is the user's knowledge of the subject area that is the focus or topic of the search (Wildemuth, 2004). Users with a high level of domain knowledge find more information quicker than others (Downing et al., 2005), while users with incorrect or imprecise domain knowledge might find more irrelevant information (Keselman et al., 2008). Our results showed that domain knowledge influenced users' success in information seeking. Users with a higher level of topic knowledge achieved their information needs faster by choosing the correct term and narrowing down to sub-terms and following the links under them. Some users changed the chosen terms or sub-term if they did not achieve their search goals.

The system's low responsiveness was a major concern in the visual search. Even though the visual interface made the information seeking easy, several times clicking on different terms from broad to specific was more time-consuming than entering the query in the search box and hitting enter in the basic search. Some users were frustrated with their task, as the visual interface did not have the functionality they expected. Wu et al. (2008) noted that an interface that is rated as easy to use might not be rated as satisfactory. Also, a slow response time is common when browsing the interface of a multi-faceted categorization system (Wildemuth, 2004).

In Fagan's (2006) study, users of different groups pointed out the strengths and weaknesses of basic and visual search. This study indicated that the narrowing down feature of results is a weakness of visual search, while other databases yielded broader ranges of results. Participants in Fagan's (2006) study described numerous strengths for the visual search, including an improved user interface for search, having the impression of getting results more quickly through colour instead of keywords, providing a general list of results and breaking it down to specific results. Although the participants in Fagan's (2006) study noted quicker access to results as a strength, the present study indicated opposite results.

Some participants preferred other databases regardless of the strengths of the visual search. This raised the insight that previous user search preferences would influence their current or future search behaviours as well as their interaction with the system. User familiarity with the visual search affected the simplicity or difficulty of their interaction with the interface. Several previous studies have emphasised the importance of user habits in information seeking. For instance, information-seeking habits of physicians might determine the nature of their information resource preferences (Dawes and Sampson, 2003). In the present study, habits were related to whether users were used to doing search in a visual search interface or not.

The data analysis did not yield any code indicating the importance of the time in performing search, while it has been shown to be part of the pre-requisites of other searches (Savolainen, 2006). Nevertheless, some factors were somewhat related to time, such as a system's responsiveness or accessibility of the results. Savolainen (2006, p. 110) considered time in

information seeking with three approaches:

(1) time as a fundamental attribute of situation or context of information seeking;

(2) time as qualifier of access to information; and

(3) time as an indicator of the information-seeking process.

Time has also been mentioned as a primary context while studying collaborative information seeking (Shah and González-Ibáñez, 2012). Shah and González-Ibáñez (2012) indicated that the main complaint by many participants was the time limit, as they thought they could have performed better searches if they had had more time.

The study's results showed that language and non-open-access articles could be a restriction for users. Non-open-access articles, which were not available through the institution's subscription frustrated users. Some studies have reported that language plays a significant role in the user's information-seeking strategies (Sabbar and Xie, 2016), specifically those who rely on sources that are not in their own language.

Although visual search interfaces are supposed to assist users with information seeking (Fagan, 2006), and several other studies have emphasised that bringing visualisation in search engines has been unsuccessful, unprofitable or ignored by users (Pajić, 2014). Thus, further studies are required to assess the usability of visual search interfaces. Further studies should also explore whether or not user interaction with the visual search interface would improve the ease and speed of accessing information.

**Limitations**

EBSCOhost has recently ceased using visualisation techniques for information searching. However, at the time of conducting this study, EBSCOhost was one of the only scientific databases that provided the context for studying every aspect of user interaction with visual search. Nevertheless, the current study provides us with the contextual factors influencing information seeking in a visual information system. Thus, the results could be helpful for the development of similar information systems, which use visual search functions or may provide users with such features.

**Conclusion**

This study presents the contextual factors affecting information seeking in a visual search interface, including general characteristics of the users, general search features, visual search features, display of results, accessibility of results, task type and environmental factors.

This study contributes to a better understanding of how people conduct searches in and interact with visual search interfaces. Results have important implications for the designers of IR systems.